\documentclass[twocolumn,floatfix,pra,aps,showpacs,superscriptaddress]{revtex4}
\usepackage{epsfig,graphicx,graphics}
\usepackage{flafter}
\usepackage{amsmath}
\usepackage{color}
\usepackage{bm}
\usepackage{amssymb}
\usepackage{epstopdf}
\usepackage{amstext}
\usepackage{amsthm}
\usepackage{amsfonts}
\usepackage{latexsym}
\usepackage{multirow}
\usepackage{array}

\newcommand{\da}{^\dagger}
\newcommand{\ket}[1]{\left\vert#1\right\rangle}
\newcommand{\bra}[1]{\left\langle#1\right\vert}

\renewcommand{\emph}[1]{{\it #1}}
\renewcommand{\vec}[1]{\boldsymbol{#1}}

\begin{document}

\title{Mesoscopic Bose-Einstein condensates as quantum simulators} 

\author{A. Gallem\'{\i}}
\affiliation{Departament d'Estructura i Constituents de la Mat\`{e}ria, Facultat de F\'{\i}sica, Universitat de Barcelona, E--08028
Barcelona, Spain}
\author{M. Guilleumas}
\affiliation{Departament d'Estructura i Constituents de la Mat\`{e}ria, Facultat de F\'{\i}sica, Universitat de Barcelona, E--08028 Barcelona, Spain}
\affiliation{Institut de Nanoci\`encia i Nanotecnologia de la Universitat de Barcelona, IN$\,^2$UB, E--08028 Barcelona, Spain}
\author{R. Mayol}
\affiliation{Departament d'Estructura i Constituents de la Mat\`{e}ria, Facultat de F\'{\i}sica, Universitat de Barcelona, E--08028 Barcelona, Spain}
\affiliation{Institut de Nanoci\`encia i Nanotecnologia de la Universitat de Barcelona, IN$\,^2$UB, E--08028 Barcelona, Spain}
\author{A. Sanpera}
\affiliation{Instituci\'o Catalana de Recerca i Estudis Avan\c{c}ats, ICREA, E--08011 Barcelona, Spain} 
\affiliation{Departament de F\'{\i}sica, Universitat
Aut\`{o}noma de Barcelona, E--08193 Bellaterra, Spain}

\date{\today}

\begin{abstract}
Mesoscopic interacting Bose-Einstein condensates confined in a few traps display phase transitions that cannot be explained with a mean field theory. 
By describing each trap as an effective site of a Bose-Hubbard model and using the Schwinger representation of spin operators, these systems can be mapped to spin models. 
We show that it is possible to define correlations between bosons in such a way that critical behavior is associated to the divergence of a correlation length accompanied 
by a gapless spectrum in the thermodynamic limit. The latter is now defined as the limit in which the mean field analysis becomes valid. Such description provides critical 
exponents to the associated phase transitions and encompasses the notion of universality demonstrating thus the potential use of mesoscopic Bose-Einstein condensates as 
quantum simulators of condensed matter systems.

\end{abstract}


\maketitle

\section{Introduction}

One of the paradigmatic models of quantum magnetism is the Hubbard model. It was originally proposed as a toy model to study magnetic properties of electrons 
in metals, and has become recently one of the cornerstones of quantum simulators with ultracold atoms. The seminal proposal \cite{Jaksch1998} that ultracold 
atoms confined in sufficiently deep optical lattices can be described by a Bose-Hubbard (BH) model \cite{Fisher1989}, together with the experimental 
demonstration of the superfluid-Mott insulator quantum phase transitions in such systems \cite{Greiner2002}, has triggered the field of quantum simulators 
with ultracold atoms. Since then, the scope of condensed matter phenomena that can be addressed with lattice gases has broadened enormously \cite{Bloch2008,Lewenstein2007,Lewenstein2012,Greif2013}. 
A paradigm are spin models derived from second order perturbation theory of Bose-Hubbard Hamiltonians. In 
fact, in the Mott phase of the lattice gas, atoms are actually frozen at each lattice site, but virtual tunneling between neighboring sites acts as an effective spin exchange interaction. 
The temperatures needed to achieve such models are, however, very restrictive going down to the picoKelvin regime 
since they scale as $t^{2}/U$, where $t$ denotes tunneling between nearest sites and $U$ is the atomic two-body interaction.

Here we take a different path to show that quantum magnetism can also be approached with mesoscopic Bose-Einstein condensates (BECs) confined in just few 
harmonic traps, relaxing significantly the temperature restrictions \cite{footnote1}. The description of mesoscopic ultracold BEC confined in few well potentials has been addressed 
in great detail both experimentally \cite{Morsch2001,Albiez2005,Levy2007} and theoretically, see e.g. \cite{Morsch2006} and references therein.  
In the weakly interacting regime, these mesoscopic systems can be described within a Hartree approach in which all particles share a 
common state (the condensate wavefunction). The dynamics at low temperatures can be accurately reproduced with the Gross-Pitaevskii equation (GPE). The 
simple case of a double well can be further simplified with the so-called two-mode approximation, 
which provides an excellent description of the relevant static and dynamic properties of the system with only two variables: the population imbalance and 
the phase difference between the two wells, giving rise to Josephson physics.

The mean field description, nevertheless, fails drastically when the interaction strength between atoms approaches some critical value; the ground 
state solution of the GPE breaks the symmetry of the double well potential \cite{JuliaDiaz2010a,JuliaDiaz2010b} and becomes highly unstable. In these cases, the Bogoliubov approach, 
which takes into account non-condensate modes, also shows a divergence in the number of atoms in the non-condensate modes \cite{Zin2008}. 
This reflects the presence of massive fluctuations of the particle number in the border of the transition regime. 
Indeed, it is the existence of quantum fluctuations on all length scales, the most characteristic feature of  (continuous) quantum phase transitions 
(QPTs) occurring at $T=0$. 
This behavior is commonly denoted as criticality. Approaching the QPT, in the parameter space established by the corresponding Hamiltonian, there is 
often a common behavior of some observables. This is reflected in a set of parameters called critical exponents that determine the qualitative nature of the critical behavior. Those are independent of the microscopic 
details of the system, but are rather linked to the symmetries of the emerging order. Thus, QPTs associated to different Hamiltonians that share the same 
set of critical exponents are said to belong to the same universality class. The QPT is also accompanied by the vanishing of some energy scale and the 
divergence of some length (the correlation length) which indicates the spread of correlations in the system \cite{Sachdev1999}. 

In this regime, the accurate description of these systems is a simplified BH Hamiltonian 
where each trap corresponds now to a mode or site. Here we associate a length scale to such systems and link 
critical behavior to its divergence together with a gapless spectrum in the corresponding thermodynamic limit (TL). The latter 
is defined here as the limit in which the mean field analysis becomes valid. Moreover, such description permits to determine the critical exponents of 
the corresponding phase transitions using finite size scaling (FSS). We explicitly consider other mesoscopic systems beyond the double well, e.g. dipolar 
gases in triple wells, and show that some of them share exactly the same critical exponents in the TL and fall, therefore, in the same universality class.

\section{The model}

In what it follows we consider $N$ spinless bosons trapped in few harmonic traps.  We further assume that atoms might have a dipole 
moment $\vec{d}$ which can be either magnetic or electric and that all dipoles are oriented along the same direction by the presence 
of an external strong field. Bosons interact via short range potentials but also, when present, with dipolar long range interactions that 
couple bosons in different traps. The bosonic field operators that annihilate (create) a boson at a point $\vec{r}$ are defined as 
$\hat{\psi}(\vec{r})=\sum_{i} \phi_{i}(\vec{r}) \, \hat{a}_{i}$, where as usual $\hat{a}_{i} (\,\hat{a}_{i}^{\dagger})$ are the 
bosonic annihilation (creation) operators on trap $i$ fulfilling canonical commutation relations. Under these assumptions 
the most general dipolar Bose-Hubbard (dBH) Hamiltonian reads:
\begin{align}
{\hat H}=&-\frac{t}{2}\sum_{i}\left[\hat{a}^\dagger_{i} \hat{a}_{i+1} + h.c. \right]
+\frac{U_0}{2} \sum_{i=1}^{k} \hat{n}_i (\hat{n}_i-1) \nonumber \\
&+\sum_{i\neq j} U_{ij} \, \hat{n}_i \hat{n}_j\,,
\label{BHH}
\end{align}
where $\hat{n}_{i}=\hat{a}_{i}^{\dagger}\hat{a}_{i}$ is the particle number operator on $i$-$th$ well, 
$\sum_{i}\hat{n}_{i}=N$ and $k$ is the total number 
of wells, which we restrict to be $k\le 3$ although our study could be generalized to higher $k$. The Hamiltonian (\ref{BHH}) is 
characterized by three parameters: the tunneling rate ($t$) between adjacent wells, the on-site energy ($U_0$) which includes both 
contact and dipole-dipole interactions, and  the inter-site energy ($U_{ij}$), which takes into account the long-range and anisotropy 
of dipolar interaction. Notice, however, that for a double well potential the effects of dipolar 
interactions can be included into a rescaled contact interaction $U$ and the Hamiltonian reduces to the non-dipolar case 
\cite{Abad2011}. The structure of the BH Hamiltonian (\ref{BHH}) makes it convenient to work in the Fock basis that labels the 
number of atoms in each well: $\ket{\Psi} = \sum_{n_{i}} \, C_{n_{1},n_{2},....n_{k}} \ket {F}_{n_{1},n_{2},....n_{k}}$ where the 
Fock state $\ket F_{n_{1},n_{2},....n_{k}}=\ket{n_1}\otimes \ \cdots \otimes \ket{n_k}$, being $k$ the number of modes. We exact diagonalize 
(\ref{BHH}) for different values of $N$ and  
the phase space parameters in the vicinity of QPTs.

We first discuss the physics of the double well potential. This system can be straightforwardly 
mapped onto the Lipkin-Meshkov-Glick (LMG) spin model \cite{Lipkin1965} by using the Schwinger representation which 
maps spin operators (SU(2)-algebra) onto two harmonic oscillators, i.e. two creation (annihilation) operators 
$\hat{a}_{i}^{\da} (\hat{a}_{i}), (i=1,2)$ as $\hat{S}^{+}=\hat{a}_{1}^{\da} \hat{a}_{2}$; $\hat{S}^{-}=\hat{a}_{2}^{\da} \hat{a}_{1}$; 
and $\hat{S}_{z}=1/2(\hat{a}_{1}^{\da} \hat{a}_{1}-\hat{a}_{2}^{\da} \hat{a}_{2})$.  The holonomic constrain 
$\hat{a}_{1}^{\da}\hat{a}_{1}+\hat{a}_{2}^{\da} \hat{a}_{2}=2\hat{S}=\hat{N}$, fixes the total number of bosons or equivalently the total spin 
and cuts in this way the infinite tower of states of the harmonic oscillators. Using such representation, the two-site BH 
Hamiltonian can be rewritten as
\begin{eqnarray}
\hat{H}&=&-\frac{t}{2} (\hat{S}^{+}+\hat{S}^{-}) + U (\hat{S}_{z}^{2} +4 \hat{S}^{2}-2\hat{S})\nonumber\\
&=&-t\hat{S_{x}}+(U N)/N \,\hat{S}_{z}^{2},
\label{BBHLMG}
\end{eqnarray}
where in the last equation we have used that $[\hat{H},\hat{N}]=[\hat{H},\hat{S}]=0$ to remove all terms proportional to the total 
spin $\hat{S}$. Hamiltonian (\ref{BBHLMG}) describes a ``mean field'' Ising model, i.e. a system of spin-$1/2$ particles mutually interacting 
embedded in a transverse magnetic field along the $x$-direction, where $\hat{S}_{\alpha}=\sum_{i=1}^{N}\hat{\sigma}_{i}^{\alpha}/2$. 
The double well Hamiltonian (\ref{BBHLMG}) is just a particular case of the general LMG model 
$\hat{H}_{LMG}=-\lambda/N\,\sum_{i<j}\left(\hat{\sigma}_{i}^{z}\hat{\sigma}_{j}^{z}+\gamma\hat{\sigma}_{i}^{y}\hat{\sigma}_{j}^{y}\right)-h\sum_i\hat{\sigma}_{i}^{x}$ with $\gamma=0$, 
introduced long time ago in nuclear physics to study mean-field QPTs and since then exploited in many different contexts e.g. \cite{Botet1983,Reslen2005,Dusuel2004,Orus2008}. 
The factor $1/N$ ensures the convergence of the free energy per spin 
in the TL. For $\lambda>0$, (ferromagnetic coupling) it is well established that there exists a second order 
phase transition at $\lambda=|h|$, if $0\leq\gamma\leq 1$. 
In our double-well language, tunneling $t$ plays the role of the external magnetic field $h$ along the 
$x$-direction, while the on-site interactions act as effective spin-spin interactions. 
Thus, approaching  $|U|N/t  \rightarrow 1$ there 
is a transition between ferromagnetic and paramagnetic order, which in the limit $N\rightarrow \infty$  converges to $|U|/t \rightarrow 0$. 
The paramagnetic region is thus proportional to $1/N$ and shrinks to zero in the TL. Notice that in these models, in which each particle 
interacts with each other, the concept of length is not defined.

\section{Correlations and critical exponents}

We aim at providing a definition of correlations, which naturally embraces the notion of correlation length 
and allows to link critical behavior to its divergence. With this purpose we first calculate the phase diagram of the double well for different values of $N$ near criticality. 
Then, we analyze the scaling behavior of some operators and performing FSS we obtain the corresponding critical exponents. Inspired by two point correlations in spin chains we 
define correlations in our system whose behaviour properly displays the features of QPTs.
Finally, we check if other models of the restricted Bose-Hubbard family 
share the same critical exponents and belong, therefore, to the same universality class.  

Notice that in spin chains the length is naturally settled by the number of sites $L$, and two-body correlations are given by the 
$\mathcal{C}_{ij}=\langle \vec{S}_{i}\vec{S}_{j}\rangle-\langle \vec{S}_{i}\rangle\langle\vec{S}_{j}\rangle$. Translational invariance ensures that their behavior 
depends on the distance between the two sites $|i-j|$, but not on the specific sites $i,j$. The latter allows to define the correlation 
length $\xi$ which fixes the length scale at which all spins are correlated between them. Far from criticality the decay is exponential, $\mathcal{C}_{ij}\sim \exp(-|i-j|/\xi)$.  
At criticality, for continuous second order phase transitions, the decay is algebraic  $\mathcal{C}_{ij} \sim (|i-j|^{-(d-2+\eta)})$ and the correlation length diverges as $\xi \propto |U-U_{crit}|^{-\nu}$, 
where $U_{crit}$ is the critical point.

\begin{figure}[h!]
\centering
\includegraphics[width=0.97\linewidth, clip=true]{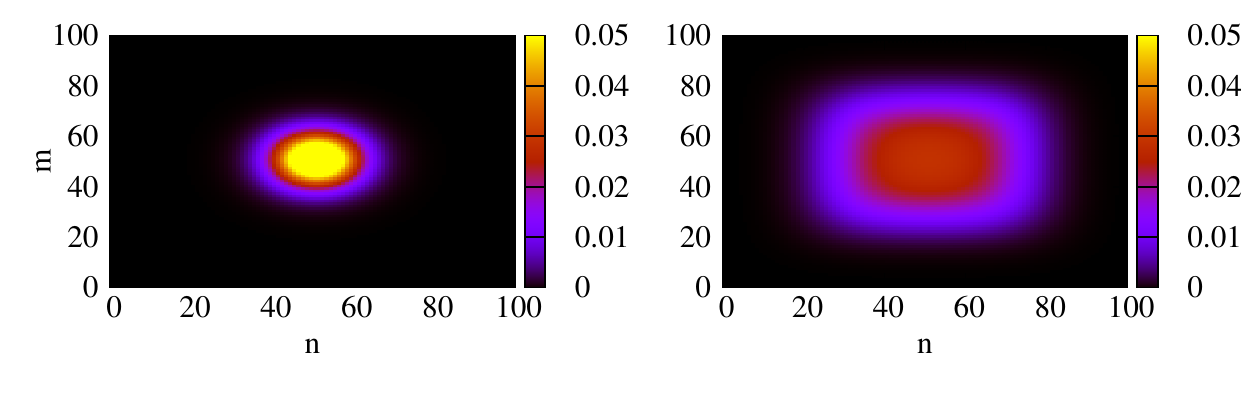}
\caption{Behaviour of $\mathcal{C}_{nm}$ as a function of $n$ and $m$. Left panel for $UN/t=60$, far from criticality; right panel, for $UN/t=-0.996$, at criticality.}
\label{Fig1}
\end{figure}

\begin{figure}[h!]
\centering
\includegraphics[width=0.48\linewidth, clip=true]{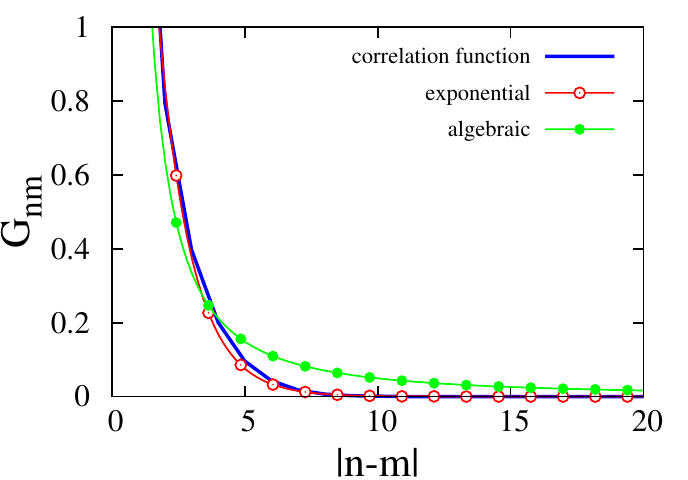}
\includegraphics[width=0.48\linewidth, clip=true]{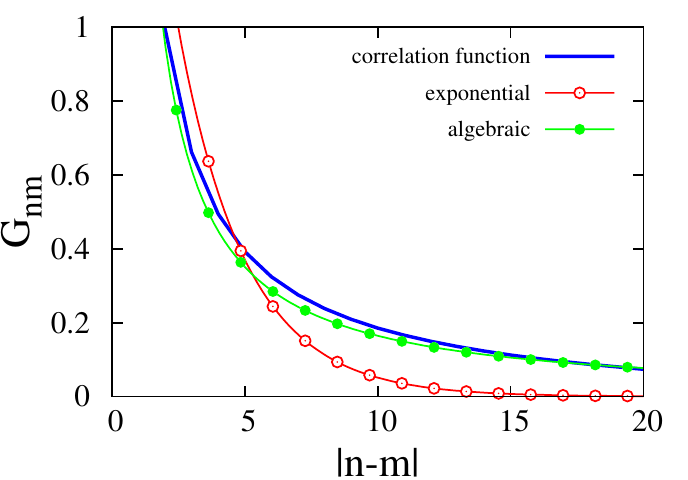}
\caption{Left: $G_{nm}$ versus distance far from criticality (left) and at criticality (right). We have fitted $G_{nm}$ by an exponential and an 
algebraic decay.  At criticality, the best adjustment is algebraic, otherwise, the best is the exponential one. The values of $UN/t$ are the same as in Fig. \ref{Fig1}.
}
\label{Fig2}
\end{figure}

In a double well, such length scale is obviously irrelevant. In order to mimic the behavior of second order QPTs in 
spin chains, we first associate ``length'' to the number of bosons, since this is the quantity that settles the dimension of the 
corresponding Hilbert space. Furthermore, this permits to order the Fock states 
in the following way: $|N,0\rangle,|N-1,1\rangle,\cdots,|0,N\rangle$. We define two-body correlations in our system as: 
\begin{equation}
 G_{nm}=\frac{\ket{n}\bra{m} \otimes \ket{ N-n}\bra{N-m}}{|n-m|}= \frac{\mathcal{C}_{nm}}{|n-m|},
 \label{correfunc}
\end{equation}
where the operator $\ket{n}\bra{m}$ acts on the first trap and $\ket{ N-n}\bra{N-m}$ on the second one.
Notice that $\mathcal{C}_{nm}$  is not equivalent to the widely used population imbalance. The latter indicates the difference on population between the left and right wells and corresponds to the expectation value of 
$\langle \hat{S}_{z}\rangle=\langle |\hat{n}_{1}-\hat{n}_{2}|\rangle/N=\sum_{n_{1},n_{2}}|C_{n_{1},n_{2}}|^{2}(|n_{1}-n_{2}|)/N$, while the former, 
$\mathcal{C}_{nm}=C_{n,N-n}C_{m,N-m}$, correlates boson occupation numbers within a well. In order to recover the 
``translational invariance'' concept rooted in spin chains, we weight the correlation function
by the ``effective distance'' $|n-m|$. This renormalization factor ensures the proper behavior of correlations as it neglects 
contributions from any intermediate level between $n$ and $m$. 
The general behavior of the correlations $\mathcal{C}_{nm}$, far from and close to criticality, is shown in Fig. \ref{Fig1}.
While  the spreading of correlations at criticality is already evident in $\mathcal{C}_{nm}$,
it is indeed the weighted function $G_{nm}$ the one which allows to recover the concept of
``translational invariance''  and define 
a correlation length $\xi$ for the system. In Fig. \ref{Fig2}, we display $G_{nm}$ as a function of $|n-m|$ far from and near
criticality.  As can be seen, $G_{nm}$ decays exponentially in the former case and  algebraically in the latter.  

\begin{figure}[h]
\centering
\includegraphics[width=0.48\linewidth, clip=true]{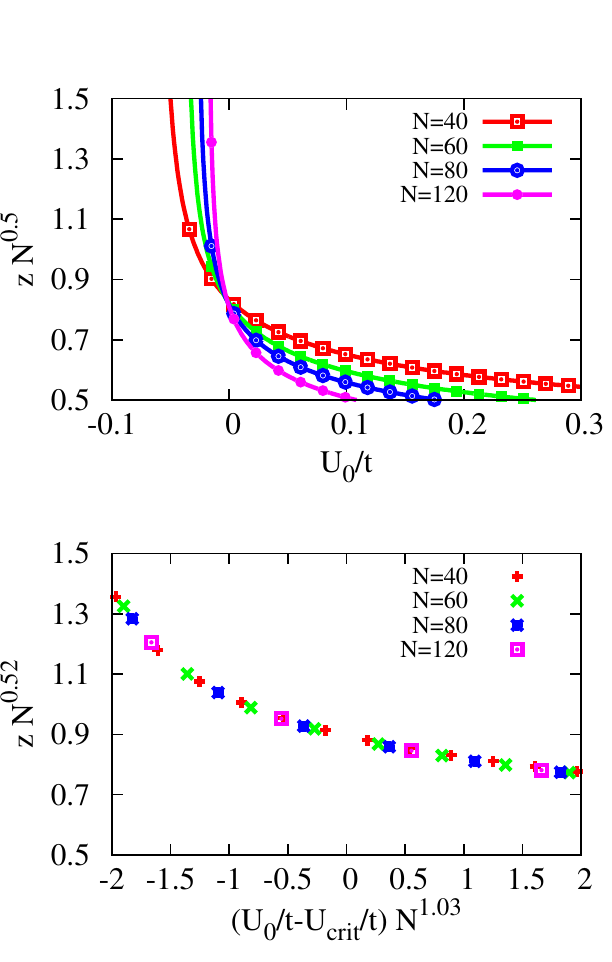}
\includegraphics[width=0.48\linewidth, clip=true]{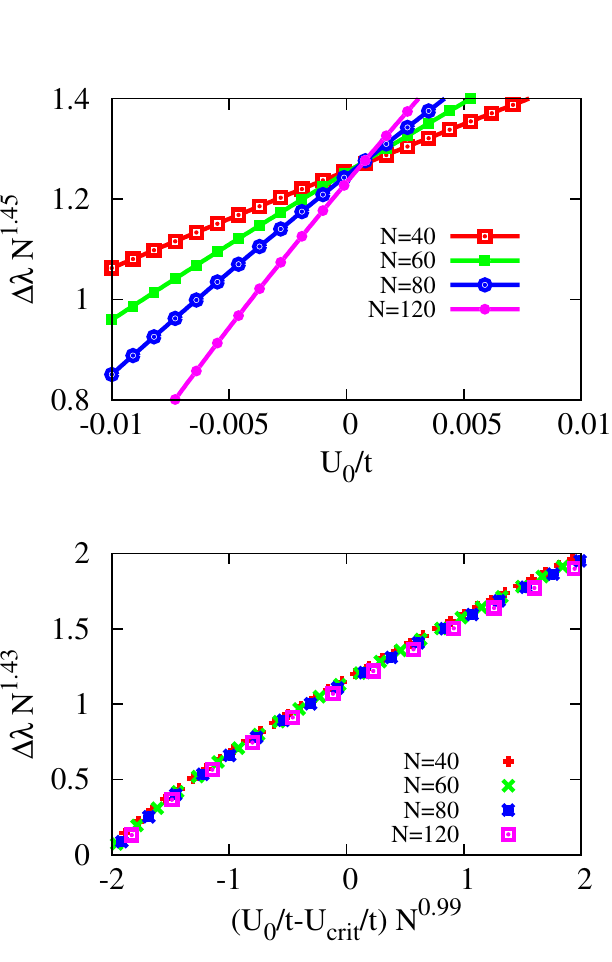}
\caption{Scaling behaviour of the population imbalance (left) and the Schmidt gap (right) in the double well potential. The critical exponents obtained via this 
method are summarized in Table \ref{tab}.}
\label{Fig3}
\end{figure}

To obtain a deeper understanding of criticality in this system as well as the exact location of the phase transition in the mean field limit, 
we analyze the population imbalance $\hat{z}$, its fluctuations and the entanglement spectrum. The latter is defined as the eigenvalues of the 
reduced density matrix of one mode or trap $\hat{\rho}_{L}=\mbox{Tr}_{R}\ket{\Psi}\bra{\Psi}=\sum\lambda_{i}\ket{u_{i}}\bra{u_{i}}_{L}$ 
where $L(R)$ stands for the left (right) trap in the double well configuration. In spin chains it has been demonstrated that the Schmidt gap, 
defined as the difference between the two largest non-degenerate eigenvalues of the entanglement spectrum (Schmidt eigenvalues), 
$\Delta\lambda=\lambda_{1}-\lambda_{2}$, closes at the critical point in the TL \cite{DeChiara2012,Lepori2013}. Finite size effects inhibit such behaviour, which can be recovered from FSS \cite{Fisher1972}.  
In Fig. \ref{Fig3} we display the scaling behavior of both, $\hat{z}$ and $\Delta\lambda$. Despite the fact that $\Delta\lambda$ is not even 
an observable, both quantities exhibit scaling i.e. they scale near a critical point as $ \hat{O} \simeq N^{\beta/\nu} f(|U-U_{crit}|N^{-1/\nu})$ 
where $N$ is now the boson number, 
$\nu$ is the mass gap exponent (associated to the correlation length divergence), and $\beta$ is the critical exponent of the corresponding operator $\hat{O}$ ($\hat{z}$ or $\Delta\lambda$). The 
clear scaling behavior of both quantities allows us to extract the critical exponents, which are summarized in Table \ref{tab}. 

It is worth mentioning that in infinitely correlated models such as the LMG, critical exponents obtained in the mean field limit, i.e. assuming a large classical spin, and 
critical exponents for finite but large $N$ are not equivalent \cite{Botet1983,Reslen2005,Dusuel2004}.  As expected, from the scaling of population imbalance (expectation value 
of the magnetization along the $z$-axis), we obtain the mean field critical values for the magnetization and mass gap: $\beta=1/2$ and $\nu=1$ \cite{Dusuel2004,Buonsante2012}. 
To further check that our definition of correlations is physically sound we compute from the algebraic decay of $G_{nm}$ (see Fig. 2 ) the exponent $(d-2+\eta)$, where $d$ is the 
dimensionality of the system. The algebraic adjustment yields the value $(d-2+\eta)=1$. Critical exponents are not all independent, they verify some algebraic relations valid in 
the TL such as $2\beta=\nu(d+\eta-2)$, which using the above expression reduces to $2\beta=\nu$. In our system, this relation is fulfilled with the mean field critical exponents. 
Interestingly enough, the critical exponents obtained from the scaling of $\Delta\lambda$ coincide with the finite size critical exponents of the LMG model $\beta=3/2$ and $\nu=1$ \cite{Botet1983,Reslen2005}. 

\begin{table}[h!]
 \begin{center}
      \begin{tabular}{| c || c | c | c | c | c | c |}
    \hline
    \multirow{2}{*}{} & \multicolumn{3}{|c}{2-well} & \multicolumn{3}{|c|}{3-well}\\
    \cline{2-7}
    & $\Delta\lambda$ & $\hat{z}$ & $\Delta \hat{z}$  & $\Delta\lambda$ (nD $\triangle$) & $\Delta\lambda$ (D $\triangle$) & $\Delta\lambda$ (D $-$) \\
    \hline
    \hline
    $\beta$ & 1.44 & 0.504 & 0.495 & 1.455 & 1.406 & 3.069 \\
    $\nu$ & 1.010 & 0.970 & 0.990 & 1.010 & 0.990 & 3.448 \\
    \hline
    \end{tabular}
 \end{center}
 \caption{Critical exponents of the correlation length ($\nu$) and the scaling operator ($\beta$) for different models and QFT. For the double-well the 
 scaling operators are: the Schmidt gap ($\Delta\lambda$), the population imbalance ($\hat{z}$) and its fluctuations ($\Delta\hat{z}$). For the 
 triple-well: the Schmidt gap in three different geometries: the non-dipolar 
triangular configuration (nD $\triangle$) \cite
{Dell'Anna2013}, the dipolar triangular configuration (D $\triangle$) \cite{Gallemi2013} and the dipolar linear configuration 
(D $-$) \cite{Lahaye2010,Peter2012}.}
\label{tab}
\end{table}

Finally, we extend our study to a triple well configuration where we introduce also dipolar interactions responsible of long range and anisotropy 
effects. We have addressed here: (i) the isotropic triangular configuration \cite{Dell'Anna2013}, (ii) the 
anisotropic (dipolar) triangular configuration \cite{Gallemi2013}, and (iii) the linear triple-well configuration with dipolar long-range effects 
\cite{Lahaye2010,Peter2012}. The corresponding phase diagrams are obtained for different values of $N$ up to $N=60$. To localize quantum phase transitions in the TL, the scaling 
behavior of the Schmidt gap is examined near criticality. We rely only on the scaling properties of this quantity 
since no obvious definition for the order parameter exists for $k>2$. A more detailed description of 
these systems together with the corresponding QPT of interest can be found in the Appendix. Our results are also summarized in Table \ref{tab}, where a close inspection shows 
that some QPTs share exactly the same critical exponents while others are clearly different. 
As it is well established in the Landau theory of phase transitions, the universality of QPTs is determined by the underlying symmetries and dimensionality of the Hamiltonian. 
Here we find that parity is the broken symmetry in all QPTs that have the same critical exponents as the double well. (The details of the corresponding QPTs are provided in the 
Appendix). As a consequence they all fall in the same universality class of the mean field Ising spin models.

\section{Conclusions}

To summarize, we have shown that mesoscopic interacting Bose-Einstein condensates displaying quantum phase transitions can be mapped to spin models. Guided by spin chains, we have proposed a sound definition of two body quantum 
correlations in these systems which incorporates the concept of length and ``translational invariance''. This allows  to demonstrate that mean field QPTs can be also associated 
to the divergence of a correlation length which yields the mass gap exponent. By scaling arguments we have obtained critical exponents and checked that those obey the well known 
algebraic relations \cite{Mussardo2010}. Finally, we have analyzed the meaning of universality in these systems. By studying different QPTs we have verified that some of them share 
the same critical exponents and fall, therefore, in the same universality class. These facts strongly support the suitability of mesoscopic BECs as quantum simulators of condensed matter.

\acknowledgements
We acknowledge financial support from the Spanish MINECO (FIS2008-01236 and FIS2011-28617-C02-01) and the European Regional development Fund, Generalitat de Catalunya 
Grant No. SGR2014-401 and SGR2014-946. A.G. is supported by Generalitat de Catalunya Grant FI-DGR 2014. We thank useful discussions with G. De Chiara, S. Campbell and M. Moreno-Cardoner.

 \appendix

 \section{Quantum phase transitions of the triple-well systems}

\begin{center}
\begin{figure}
\begin{tabular}[t]{ m{16mm} r  m{11cm}  }
& a) &
\includegraphics[width=0.4\linewidth, clip=true]{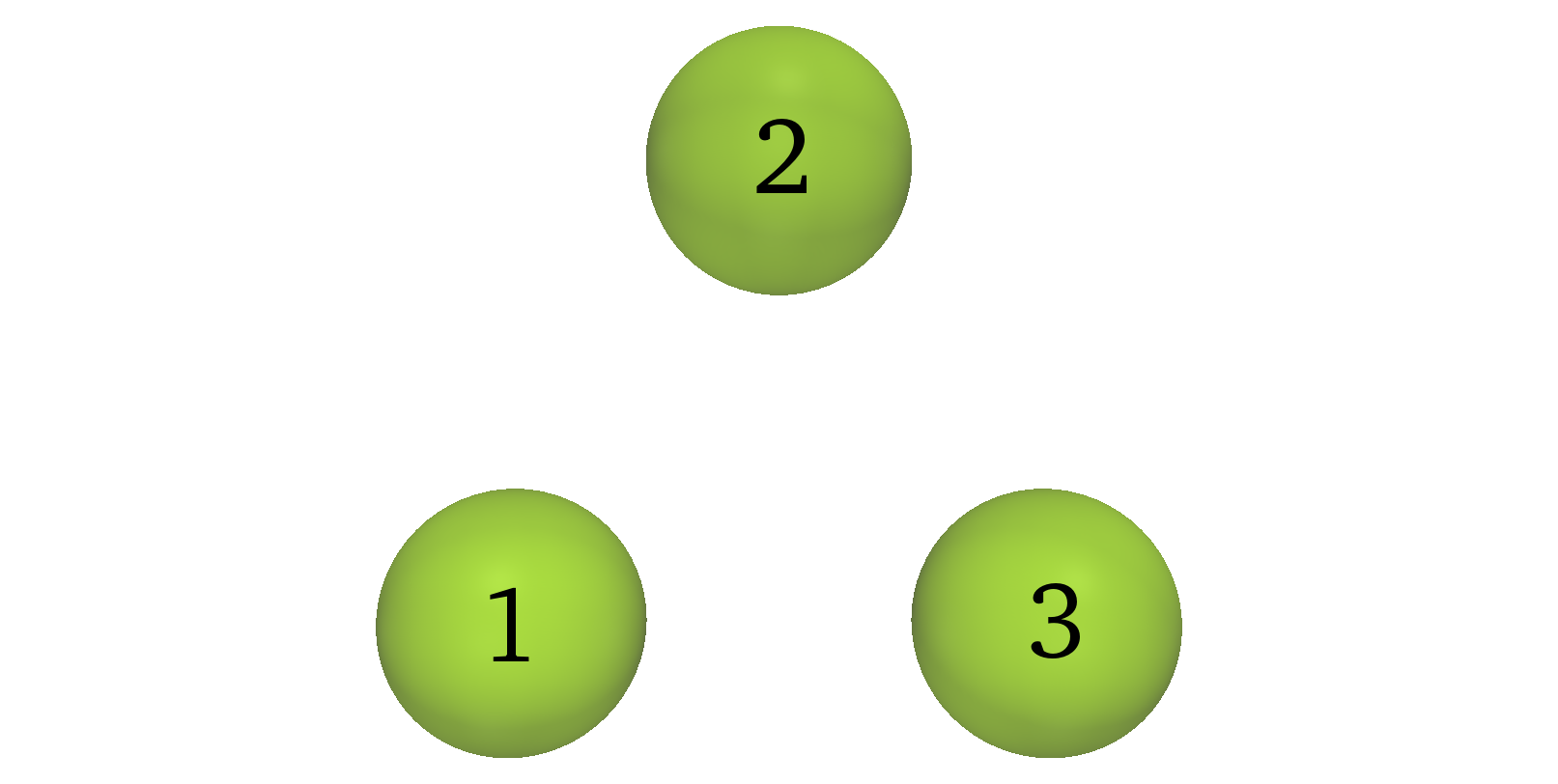}
\\
& & \\
& b) &
\includegraphics[width=0.4\linewidth, clip=true]{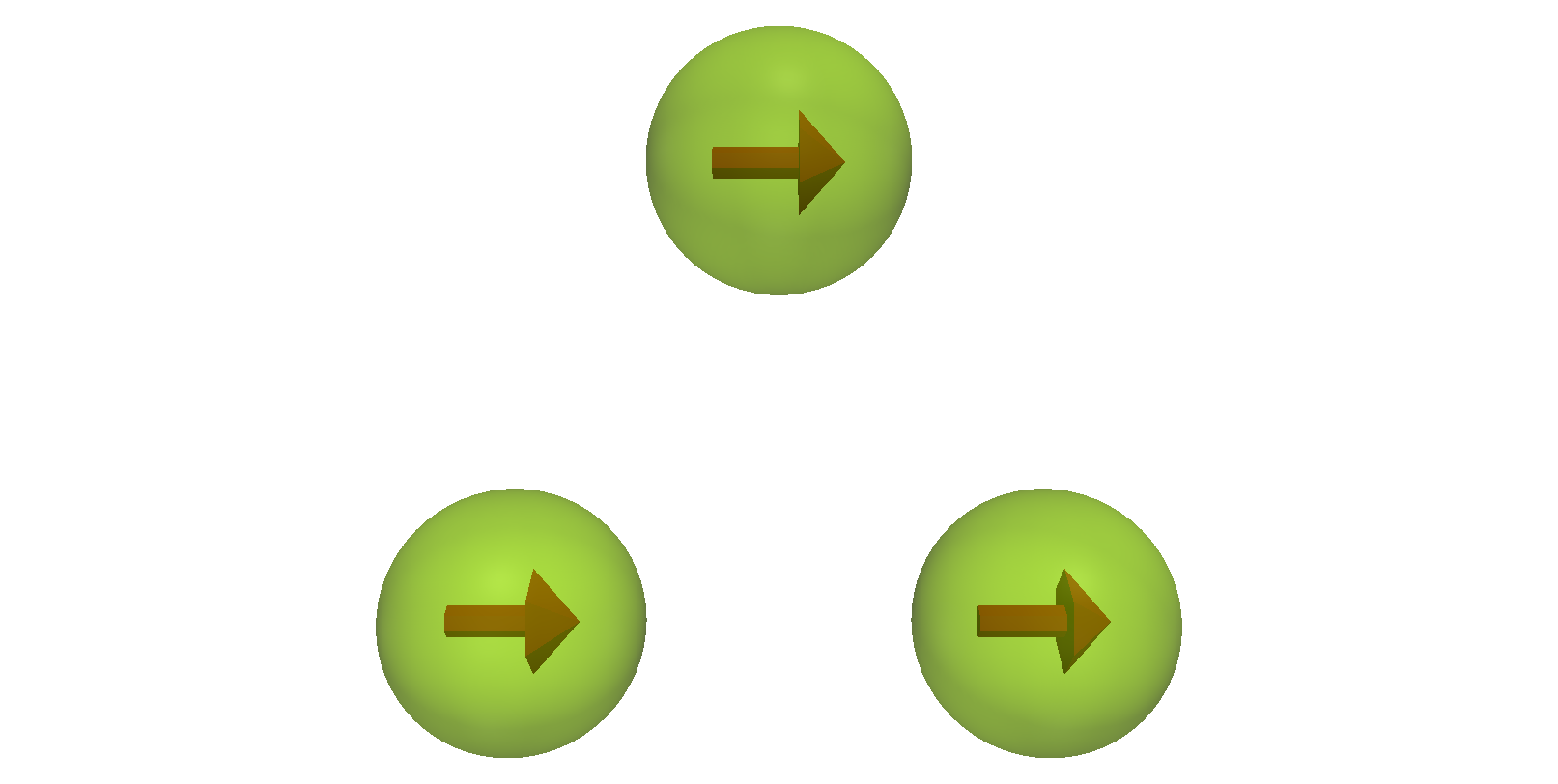}
\\
& & \\
& c) &
\includegraphics[width=0.4\linewidth, clip=true]{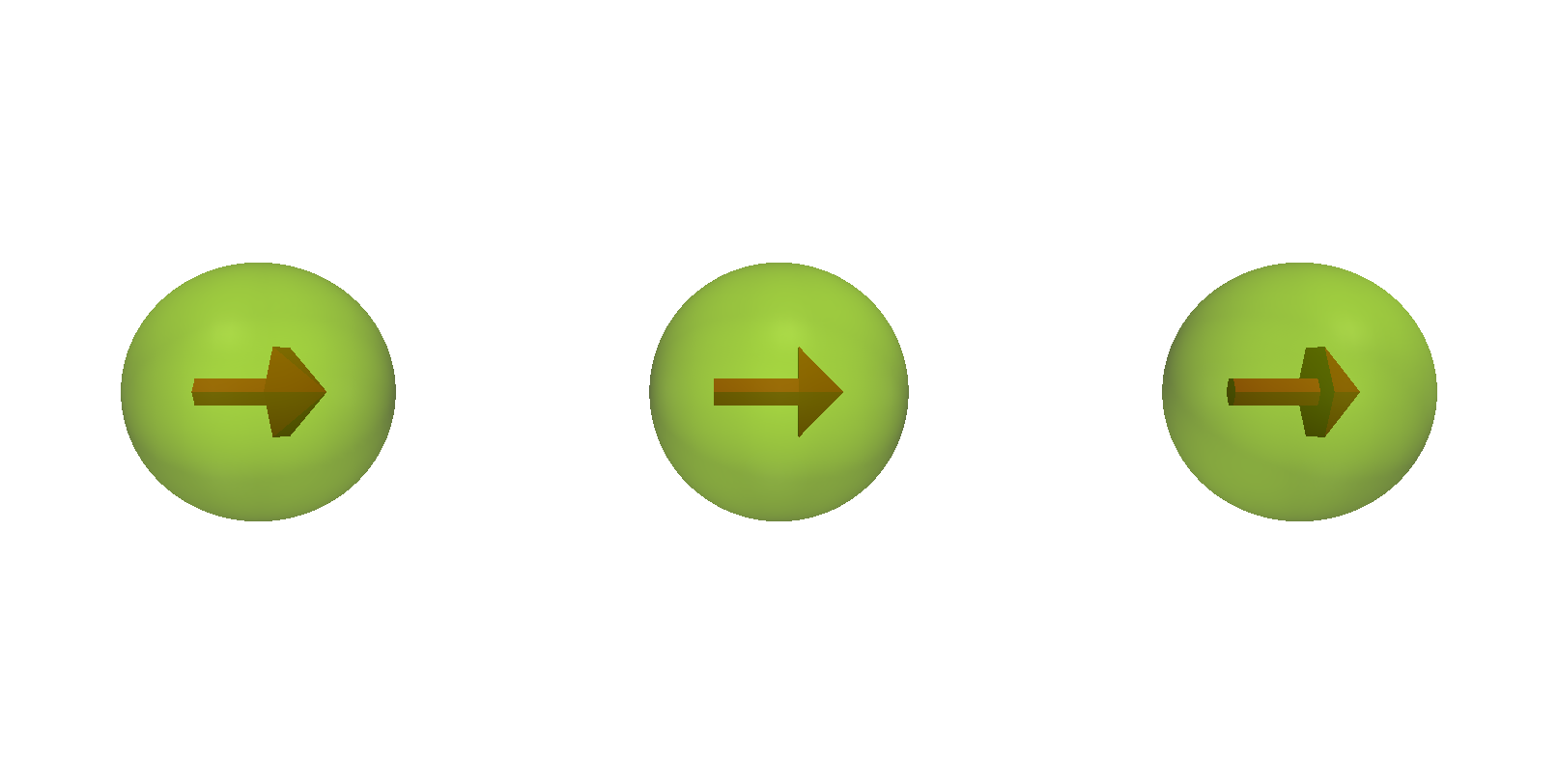}
\\
\end{tabular}
\caption{Different triple-well configurations: a) triangular non-dipolar, b) triangular dipolar c) linear dipolar. The labels $1$,$2$ and $3$ are the ones used in the text.}
\label{Fig4}
\end{figure}
\end{center}

In the main paper we have studied four different systems concerning to different few lattice site configurations. In the simplest one, which is the double-well configuration, 
scaling properties have been studied by means of the entanglement (Schmidt gap) as well as the ``magnetization'' (population imbalance) properties.  By extending the system to three wells, 
we introduce long range and anisotropic interactions resulting from the dipolar interactions. 
The triangular non dipolar case \cite{Dell'Anna2013}, the dipolar one \cite{Gallemi2013} and 
the linear geometry with dipolar interaction \cite{Lahaye2010} are the particular cases studied in this work, see Fig. \ref{Fig4}. In this section we present how are their 
phase diagrams in the vicinity of the quantum phase transition we focus on in our work. 

\begin{figure*}[t!]
\centering
\includegraphics[width=0.32\linewidth, clip=true]{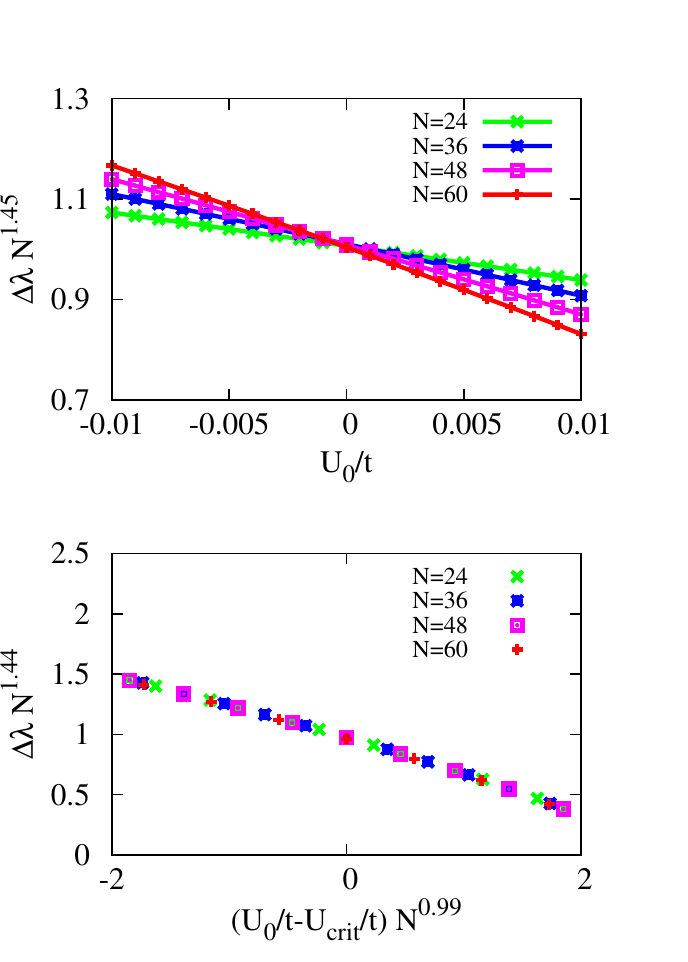}
\includegraphics[width=0.32\linewidth, clip=true]{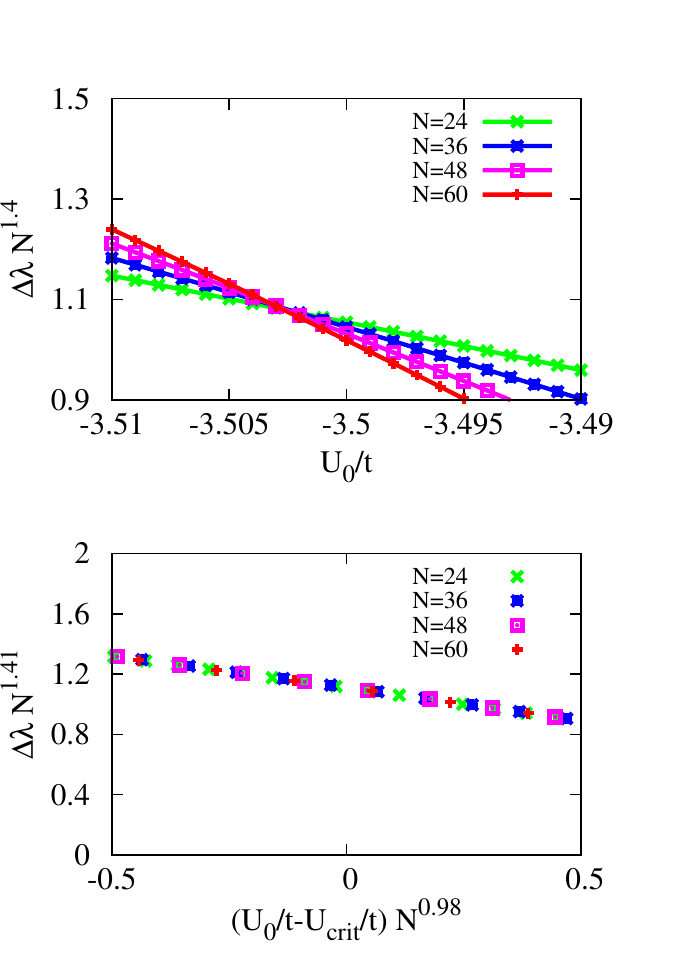}
\includegraphics[width=0.32\linewidth, clip=true]{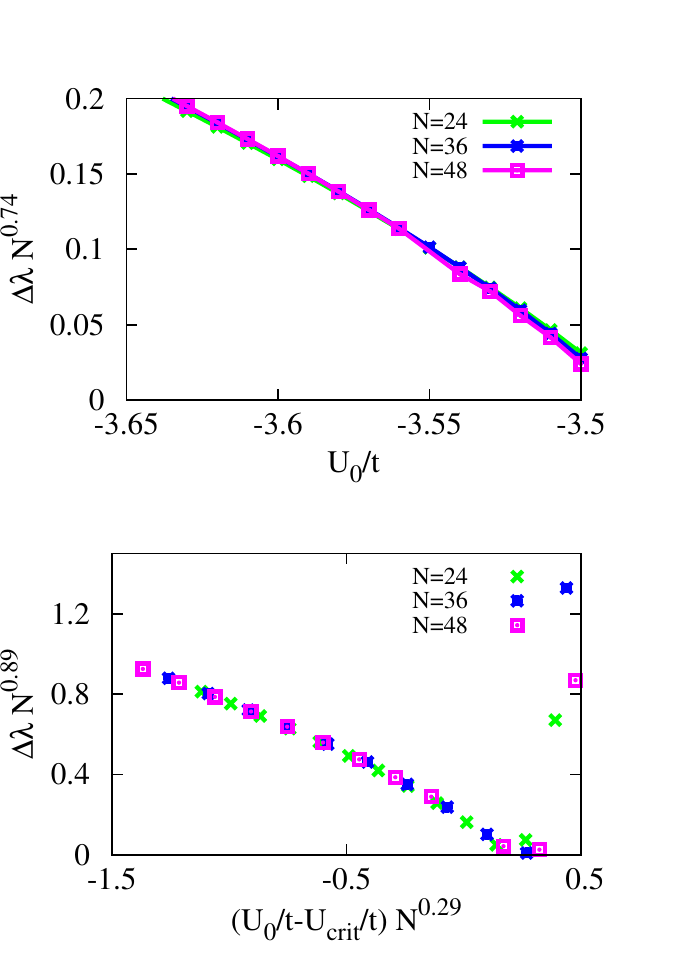}

\caption{FSS of the Schmidt gap in the different triple-well configurations. Left: triangular non-dipolar, middle: triangular dipolar, right: 
linear dipolar.}
\label{Fig5}
\end{figure*}

The non-dipolar triangular triple-well configuration a) is the most trivial extension of the double-well \cite{Dell'Anna2013}, due to the fact that the rotational symmetry is still 
preserved. Two phases appear, and the transition between both is the one studied in the paper. The first, for repulsive interactions, essentially consists in an equipartition of the particles 
among the three wells $|\Psi \rangle \approx |N/3,N/3,N/3\rangle$. When the interaction is switched to be negative (attractive), a W-state solution is encountered when diagonalizing the Hamiltonian, i.e. 
$|\Psi \rangle \approx 1/\sqrt{3} (|N,0,0\rangle+|0,N,0\rangle+|0,0,N\rangle)$.

\begin{figure}[h!]
\centering
\includegraphics[width=0.6\linewidth, clip=true]{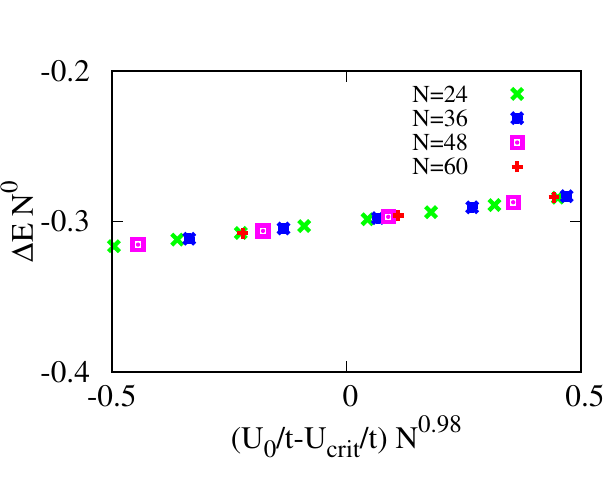}
\includegraphics[width=0.6\linewidth, clip=true]{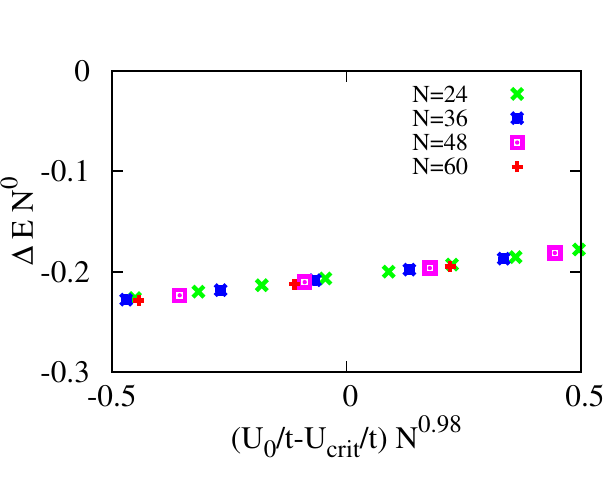}

\caption{FSS of the mass gap in the different triangular triple-well configurations. Top: non-dipolar. Bottom: dipolar.}
\label{Fig6}
\end{figure}

The dipolar case b) is noticeably more complex. The rotational symmetry disappear, nevertheless, inversion symmetry $L/R$ still manifests in the system. 
Many phases appear; however, in our work, we study only the transition between two phases, labelled D and E in Ref. \cite{Gallemi2013}. 
The former is an equipartition of the particles between wells 1 and 3, $|\Psi \rangle \approx |N/2,0,N/2\rangle$, whereas the latter is a cat state between wells 1 and 3, 
$|\Psi \rangle \approx 1/\sqrt{2} (|N,0,0\rangle+|0,0,N\rangle)$. Since well 2 is completely empty in this region of the phase diagram, the system behaves as an effective double-well system.

Finally, the system concerning dipolar bosons loaded on an aligned triple-well potential c) is characterized by four regions, see Ref. \cite{Lahaye2010}. In this case, the ``C-D'' 
transition is studied. In phase C, all the particles are in the central well, $|\Psi \rangle \approx |0,N,0\rangle$, while phase D corresponds to a cat state between the external wells 
with half of the particles in the central well, $|\Psi \rangle \approx 1/\sqrt{2} (|N/2,N/2,0\rangle+|0,N/2,N/2\rangle)= 1/\sqrt{2}\ket{N/2}_{2}(\ket{N/2}_1\ket{0}_{3}+
\ket{0}_1\ket{N/2}_{3})$.

\section{Finite Size Scaling for the triple-well configuration}

In the different triple-well configurations we have performed Finite Size Scaling (FSS) for two different quantities: the mass gap (energy gap between the 
ground state and the first excited state), and the Schmidt gap of the partition $1/23$. The reason why we have chosen such partition instead of any other is because 
this choice guarantees not to lose information by symmetry arguments. Fig. \ref{Fig5} shows the FSS analysis of the Schmidt gap for the different triple-well 
configurations. The corresponding critical exponents are summarized in the table of the main paper. In order to verify our results, we also calculate the 
critical exponent associated to the mass gap, that is, the energy difference between the first excited state and the ground state. It must scale with an exponent  
$\Delta E \propto |U /t - U_{crit}/t|^{\nu}$, where $\nu$ corresponds to the critical exponent obtained from the FSS analysis of other magnitudes. From Fig. \ref{Fig6} 
we obtain $\nu=1$ in agreement with the double well potential.

\end{document}